\documentclass[11pt]{article}
\pdfoutput=1  
\usepackage{graphicx}

\usepackage{appendix}

\usepackage{amssymb}
\usepackage{amsmath}
\usepackage{amsfonts}
\usepackage{upgreek}
\usepackage{latexsym}
\usepackage{slashed,upgreek,amscd,cancel,tensor}
\usepackage{adjustbox}
\usepackage{epsfig}
\usepackage{cite}
\usepackage{hyperref}
\usepackage{amsfonts}
\usepackage{booktabs}
\usepackage{siunitx}
\usepackage{caption}
\usepackage{float}
\usepackage{color}

\numberwithin{equation}{section}

\definecolor{MyBlue}{rgb}{0.15,0.15,0.70}

\hypersetup{
colorlinks=true,
citecolor=MyBlue,
linkcolor=MyBlue,
urlcolor=MyBlue
}

\input epsf
\newcommand{\be}{\begin{equation}}
\newcommand{\ee}{\end{equation}}
\newcommand{\beq}{\begin{equation}}
\newcommand{\eeq}{\end{equation}}
\newcommand{\bea}{\begin{eqnarray}}
\newcommand{\eea}{\end{eqnarray}}

\newcommand{\R}{R}

\usepackage[stable]{footmisc}

\def\dkmu2{\delta K_{\mu \nu}\delta K^{\mu \nu}}
\def\pmu2{  \phi_{\mu \nu}\phi^{\mu \nu}}
\def\stackrel#1#2{\mathrel{\mathop{#2}\limits^{#1}}}

\renewcommand\[{\left[}

\newcommand\ees{\end{eqnarray}}
\newcommand\bees{\begin{eqnarray}}

\newcommand\alphaB{\alpha_{\text{B}}}

\newcommand\alphaK{\alpha_{\text{K}}}
\newcommand\alphaT{\alpha_{\text{T}}}
\newcommand\alphaH{\alpha_{\text{H}}}

\newcommand{\quadac}{{\rm  quad}}

\newcommand{\2}{{(2)}}

\newcommand{\aL}{\alpha_{\rm L}}

\newcommand{\bun}{\beta_1}
\newcommand{\bdeux}{\beta_2}
\newcommand{\btrois}{\beta_3}

\newcommand{\CI}{{\cal C}_{\rm I}}
\newcommand{\CII}{{\cal C}_{\rm II}}

\newcommand{\Xt}{\tilde{X}}
\newcommand{\gt}{\tilde{g}}

\begin{document}

\begin{center}
\LARGE{\bf Mimetic gravity as DHOST theories}
\\[1cm] 

\large{David  Langlois$^{\rm a}$,  Michele Mancarella$^{\rm b}$, Karim Noui$^{{\rm c},{\rm a}}$, Filippo Vernizzi$^{\rm b}$}
\\[0.5cm]

\small{
\textit{$^{\rm a}$
  Astroparticule et Cosmologie, Universit\'e Denis Diderot Paris 7, CNRS \\
10 Rue Alice Domon et L\'eonie Duquet 75013 Paris, France}}
\vspace{.2cm}

\small{
\textit{$^{\rm b}$ Institut de physique th\' eorique, Universit\'e  Paris Saclay, CEA, CNRS \\ [0.05cm]
 91191 Gif-sur-Yvette, France}}

\vspace{.2cm}

\small{
\textit{$^{\rm c}$ Institut Denis Poisson,
Universit\'e d'Orl\'eans, Universit\'e de Tours, CNRS, 
\\[0.05cm]
Parc de Grandmont, 37200 Tours, France}}

\end{center}

\begin{abstract}
We show that theories of mimetic gravity can be viewed as  
degenerate higher-order scalar-tensor (DHOST) 
theories 
that
admit an extra local (gauge) symmetry in addition to the usual diffeomorphism invariance. 
 We reformulate and classify mimetic theories in this perspective.
Using the effective theory of dark energy, recently extended to include DHOST theories, we then investigate the linear perturbations about a homogeneous and isotropic background for all mimetic theories. We also include matter, in the form of a $k$-essence  scalar field, and we derive the quadratic action for linear perturbations in this case.  
\end{abstract}

\newpage
\tableofcontents
\vspace{0.5cm}

\section{Introduction}
\label{sec1}

Mimetic Matter was introduced by Chamseddine and Mukhanov in \cite{Chamseddine:2013kea} as a model of modified gravity that  mimics cold dark matter
\cite{Capela:2014xta,Mirzagholi:2014ifa,Ramazanov:2015pha,Babichev:2016jzg}.
The original proposal was then extended to inflation, dark energy and also theories with non-singular cosmological and black hole solutions \cite{Chamseddine:2014vna,Chamseddine:2016uef,Chamseddine:2016ktu,Liu:2017puc,BenAchour:2017ivq}. 
Mimetic theories have also been studied in~\cite{Deruelle:2014zza,Arroja:2015wpa,Hammer:2015pcx,Arroja:2015yvd}. More specifically, their linear stability has been considered in \cite{Ramazanov:2016xhp,Ijjas:2016pad,Firouzjahi:2017txv,Hirano:2017zox,Zheng:2017qfs,Gorji:2017cai,Ganz:2018mqi}.
 See also e.g.~\cite{Sebastiani:2016ras} for a review on mimetic gravity
  and \cite{Horava:2009uw,Mukohyama:2009mz,Creminelli:2009mu,Blas:2010hb,Lim:2010yk} for earlier related works.

The goal of this paper is to  revisit mimetic gravity in the context of Degenerate Higher-Order Scalar-Tensor (DHOST) theories, introduced in \cite{Langlois:2015cwa} and further explored in  \cite{Langlois:2015skt,Achour:2016rkg,Crisostomi:2016czh,deRham:2016wji,BenAchour:2016fzp}, as summarized in e.g. \cite{Langlois:2017mdk} (see also \cite{Crisostomi:2017lbg,Langlois:2017dyl,Dima:2017pwp} for a study of the screening mechanism in these theories). DHOST theories are scalar-tensor theories that encompass Horndeski~\cite{Horndeski:1974wa} and  so-called Beyond Horndeski (or GLPV) theories~\cite{Gleyzes:2014dya,Gleyzes:2014qga} (another particular subclass of DHOST theories was also found earlier  in \cite{Zumalacarregui:2013pma} via disformal transformations of Einstein gravity). 
Despite the presence of second derivatives of the scalar field in the Lagrangian and higher order Euler-Lagrange equations, DHOST theories contain at most three degrees of freedom (one scalar and  two tensorial modes),
because their  Lagrangian is degenerate \cite{Langlois:2015cwa}.

Mimetic theories can be reformulated as scalar-tensor theories with second derivatives of the scalar field in their Lagrangians. Moreover, the conformal (or, more generally, the disformal) symmetry that characterizes the resulting scalar-tensor Lagrangian guarantees that the latter is degenerate, thus implying that they  form a subclass of DHOST theories, as already 
pointed out
 in \cite{Achour:2016rkg}. When mimetic theories are restricted to a quadratic or cubic dependence on the second derivatives of the scalar field, one can use the full classification of quadratic and cubic DHOST theories obtained in \cite{BenAchour:2016fzp} and identify the 
 subclasses that contain
  these mimetic theories. Mimetic theories with a quartic or higher dependence on  second derivatives of the scalar field provide 
  examples of DHOST theories that are not included in the classification of \cite{BenAchour:2016fzp}.

In the present work, we draw upon our recent analysis of DHOST theories \cite{Langlois:2017mxy}, based on the effective approach to dark energy developed in \cite{Gubitosi:2012hu,Gleyzes:2013ooa,Gleyzes:2014rba}, to study linear perturbations in  
theories of mimetic gravity  using their DHOST formulation. 
Our calculations are consistent with those 
presented 
in \cite{Takahashi:2017pje}, based on the Lagrange multiplier formulation.

The paper is structured as follows. In Section \ref{sec2}, we start with a brief introduction to
mimetic gravity, presenting  two equivalent formulations: 
the DHOST formulation, which we exploit in this work, and the Lagrange multiplier formulation, which has often been used in the literature.  
We also show that mimetic theories, with a quadratic or cubic dependence on  second derivatives of the scalar field, 
belong
to specific subclasses of DHOST theories.
We then turn to the study of linear cosmological perturbations and first  review  the effective theory  of dark energy,  in Section \ref{Sec:EFT}. Concentrating on the DHOST formulation of mimetic gravity, we obtain   the quadratic action for  linear cosmological perturbations. 
For completeness, we briefly present  the study of  linear perturbations  within this alternative  formulation in Section \ref{Lagrangemult}. 
 We summarize our work and conclude in the final section. A few technical details are summarized in the Appendix.

\section{Mimetic gravity: short review and classification}
\label{sec2}

In this section, we first give a brief review of mimetic gravity before
constructing  the general mimetic gravity action which propagates at most three degrees of freedom. 

\subsection{Non-invertible disformal transformation}
Mimetic gravity is a scalar-tensor theory defined by a general action of the form
\bea\label{mimetic}
S[\gt_{\mu\nu},\phi] = \int d^4x \, \sqrt{ -g} \, 
{\cal L}(\phi,\partial_\mu \phi, \nabla_{\mu}\!\nabla_\nu \phi\, ; g_{\mu\nu}) \, ,
\eea
where the variation must be taken with respect to $\phi$ and the auxiliary metric $\gt_{\mu\nu}$, related to $g_{\mu\nu}$ via a  {\it non-invertible} disformal transformation,
\bea\label{disformal}
g_{\mu\nu} \; = \; \tilde A(\phi,\Xt) \, \gt_{\mu\nu} \, + \, \tilde B(\phi,\Xt)\,  \partial_\mu\phi\,  \partial_\nu\phi  \, ,\qquad \Xt \equiv \gt^{\mu\nu}\partial_\mu\phi\,  \partial_\nu\phi\,.
\eea
Note that in the original model of mimetic dark matter, the Lagrangian in \eqref{mimetic} depends only on $g_{\mu \nu}$ and not explicitly on $\phi$, since the Lagrangian is the Einstein-Hilbert term for $g_{\mu \nu}$, ${\cal L}=R$.
In the following, we will use the notation $\phi_\mu\equiv\partial_\mu\phi$  and ${\phi}_{\mu\nu} \equiv {\nabla}_\mu \phi_\nu$.

The fact that the disformal transformation is non-invertible implies that the functions 
$\tilde A$ and $\tilde B$ are not arbitrary but are related according to~\cite{Bekenstein:1992pj}
\bea\label{non-invertibility}
\frac{\tilde A(\phi,\Xt)}{\Xt} + \tilde B(\phi,\Xt)  \; = \; h(\phi) \,,
\eea
where $h(\phi)$ is an arbitrary function of  $\phi$. 
Due to the non-invertibility condition \eqref{non-invertibility}, the metric ${\gt}_{\mu\nu}$ cannot be fully determined from
$g_{\mu\nu}$ and $\phi$. Indeed, the metric $g_{\mu\nu}$ is left invariant under the local transformations
\bea\label{transfos}
\delta \phi=0 \,  \quad \text{and} \quad
\delta \tilde g_{\mu\nu} = \varepsilon \left(\tilde A_{,\tilde{X}} \, \tilde g_{\mu\nu} + \tilde B_{,\tilde X}\,  \phi_\mu \phi_\nu \right)
\eea
of the field $\phi$ and the metric $\tilde g_{\mu\nu}$, where $\varepsilon$ is an arbitrary function and 
$\tilde A_{,\tilde X}$ (or $\tilde B_{,\tilde X}$) denotes the derivative of $\tilde A$ (or $\tilde B$) with respect to the variable $\tilde X$.  
As a consequence, \eqref{transfos} defines a local invariance of 
the mimetic action. 
In general, the finite version of the infinitesimal transformation \eqref{transfos} can only be  defined implicitly. For $\tilde A = -\tilde X$ and $\tilde B=0$, as in the original mimetic theory \cite{Chamseddine:2013kea}, the finite transformation is an arbitrary conformal transformation, $\tilde g_{\mu \nu} \to C( x^\rho) \, \tilde g_{\mu \nu} $. See \cite{Horndeski:2017rtl} for a recent study of a large class of conformally invariant scalar-tensor theories, which partially overlaps with the mimetic theories studied here (as discussed in detail in the appendix of  \cite{Takahashi:2017pje}).

From a Hamiltonian point of view, the existence of this extra symmetry leads to the presence of a first-class constraint that generates 
the above gauge transformations \eqref{transfos}. This constraint adds to the usual Hamiltonian and momentum constraints associated with diffeomorphism invariance.
This has been explicitly shown in a simple case in \cite{Achour:2016rkg}.
As a consequence, mimetic gravity is necessarily a degenerate theory in the sense defined in \cite{Langlois:2015cwa} and thus belongs
to the family of  DHOST theories.

However, contrary to most  DHOST theories, the first class primary constraint in mimetic gravity does not lead to a secondary constraint, which 
is necessary to
remove the Ostrogradsky ghost, at least in the context of classical mechanics \cite{Motohashi:2016ftl,Klein:2016aiq}
 or field theory with multiple scalars \cite{Crisostomi:2017aim}. For this reason, it  was unclear whether the scalar mode in mimetic gravity is healthy or not \cite{Barvinsky:2013mea,Golovnev:2013jxa,Chaichian:2014qba}.

\subsection{Mimetic Lagrangians}
In the following, we are interested in  Lagrangians of the form
\bea\label{mimLagrangian}
 {\cal L} \; = \; {f}_2(\phi,{X}) {R} + {f}_3 (\phi,{X}) {G}^{\mu\nu} {\phi}_{\mu\nu} + 
 {\cal L}_\phi(\phi,\phi_\mu,{\phi}_{\mu\nu}) \, ,
\eea
where the only Riemann-dependent terms are proportional to  the scalar curvature $R$ and the Einstein tensor ${G}_{\mu\nu}$ 
(associated with the metric $g_{\mu\nu}$)
  to ensure that the metric carries only two tensor degrees of freedom (see e.g. discussion in \cite{BenAchour:2016fzp} and also \cite{Crisostomi:2017ugk}).
Furthermore, the relation \eqref{non-invertibility} implies immediately that ${X}$ depends only on $\phi$, since
\bea
\label{Xh}
{{X}}\equiv g^{\mu\nu}\phi_\mu\phi_\nu=\tilde A^{-1}\left(\gt^{\mu\nu}-\frac{\tilde B}{\tilde A+ \tilde B\Xt}\tilde\nabla^\mu\phi \, \tilde\nabla^\nu\phi\right)\phi_\mu\, \phi_\nu \; = \; \frac{\Xt}{\tilde A+ \tilde B\Xt} \; = \; \frac{1}{h(\phi)} \, .
\eea
Hence, the functions $f_2$ and $f_3$ depend on $\phi$ only.  As a consequence, the term proportional to $f_3$ 
can always be transformed,  via an integration by parts, into \cite{Achour:2016rkg}
\bea
{f}_3 (\phi) \, {G}^{\mu\nu} {\phi}_{\mu\nu}  \; = \; 
  \frac{{f}_{3,\phi}}{2h} {R} +  {f}_{3,\phi} [{\phi}_{\mu\nu} {\phi}^{\mu\nu} - 
 (\phi_\mu^\mu)^2] - \frac{ f_{3,\phi\phi}}{h}  \phi_\mu^\mu - \frac{h_{,\phi}}{2h^3}  f_{3,\phi\phi} \, + \, {\nabla}_\mu J^\mu \,, 
\eea
where  the explicit expression of $J^\mu$ is irrelevant here. 
For this reason, one can take 
$f_3=0$ in \eqref{mimLagrangian} without loss of generality (up to a redefinition of ${f}_2$ and ${\cal L}_\phi$).

In \cite{BenAchour:2016fzp}, only terms that  are at most cubic in second derivatives of the scalar field 
have been considered. 
Generalizing the classification of degenerate theories to higher powers of second derivatives of $\phi$ would probably  be tedious.
Interestingly, mimetic gravity provides us naturally with a 
particular 
class of DHOST theories that involves arbitrary functions of
second derivatives of $\phi$.

As shown in  App.~\ref{fieldredef} and \ref{disftran}, 
one can restrict eq.~\eqref{non-invertibility} to the case  $h(\phi)=- 1$ (for configurations with time-like gradient) and $\tilde B=0$ without loss of generality, assuming that matter is coupled to the metric $g_{\mu\nu}$. 
Therefore, for simplicity we  restrict to  the non-invertible conformal transformation 
\be
\label{conf_trans}
g_{\mu\nu}=-\tilde{X}\,  \tilde{g}_{\mu\nu} 
\ee
 in the action.
This implies the condition
\bea
\label{Xuno}
 {X} \, = \,  {g}^{\mu\nu} \phi_\mu \phi_\nu \, = \, -1 \, .
 \eea
Let us now discuss the term ${\cal L}_\phi$, involving second derivatives of the scalar field. 
This term  can be viewed as a scalar
constructed  by contracting powers of  the matrix  
\be
[{\phi}]^\mu_\nu \equiv \phi^\mu_{\ \nu} \;,
\ee
with  the vector field $\phi^\mu$ or with the metric  $g_{\mu\nu}$. Hence, it can be expressed as a function of $\phi$ and of the two sets of 
scalar quantities
$\vartheta_n \equiv \phi^\mu\phi^\nu [{\phi}]_{\mu\nu}^n$ and 
\bea
\label{chidef}
\chi_n \equiv  g^{\mu\nu}[{\phi}]^n_{\mu\nu}\,,
\eea
where $n$ is an integer. However, as $X=-1$, we have
\bea\label{nulvector}
2\, {\phi}^{\mu\nu}\phi_\nu \, = \,
{\nabla}^\mu {X} \; = \; 0 \, ,
\eea
which implies immediately that $\vartheta_n=0$ for any $n \geq 1$.
As a consequence, ${\cal L}_\phi$ is a function of $\phi$ and $\chi_n$ only, and
the mimetic action can be reduced, without  loss of generality, to the form 
\bea
\label{MimeticDHOST}
{S}[ {g}_{\mu\nu},\phi]=\int d^4 x \, \sqrt{-{g}} \left[ {f}_2(\phi) {R}  \, + \, 
{\cal L}_\phi(\phi, \chi_1, \cdots, \chi_n)  \right] \,,
\eea
where $n$ is arbitrary and where the variation is taken with respect to $\tilde g_{\mu \nu}$ related to $g_{\mu \nu}$ by eq.~\eqref{conf_trans}.

\subsection{DHOST formulation}
\label{DHOST form sec}

To illustrate the DHOST formulation of mimetic gravity, it is convenient to  work with a concrete model. 
For simplicity, we first restrict to the quadratic case
and consider 
\be
\label{quadmim}
{S}[ {g}_{\mu\nu},\phi]=\int d^4 x \, \sqrt{-{g}} \left[ {f}_2(\phi) {R}  \, + \, 
a_1(\phi)  L^\2_1  + a_2(\phi) L^\2_2  \right] \,,
\ee
where $a_1(\phi)$ and $a_2(\phi)$ are arbitrary functions of $\phi$ in front of the quadratic terms $L^\2_1 = \chi_2 $ and $L^\2_2=\chi_1^2$, and  we have used  the first two elementary quadratic Lagrangians among the five $L_A^{(2)}$ ($A=1, \ldots, 5$) introduced in \cite{Langlois:2015cwa}, i.e., 
\be
\label{QuadraticL}
\begin{split}
& L^\2_1 = \phi_{\mu \nu} \phi^{\mu \nu} \,, \qquad
L^\2_2 =(\Box \phi)^2 \,, \qquad
L_3^\2 = (\Box \phi)\,  \phi^{\mu} \phi_{\mu \nu} \phi^{\nu} \,,  \\
& L^\2_4 =\phi^{\mu} \phi_{\mu \rho} \phi^{\rho \nu} \phi_{\nu} \,, \qquad
L^\2_5= (\phi^{\mu} \phi_{\mu \nu} \phi^{\nu})^2\,.
\end{split}
\ee

Upon the change of metric \eqref{conf_trans} (see \cite{Achour:2016rkg} for details 
about  the transformations of  the quadratic action under general conformal-disformal  transformations), the action \eqref{quadmim}  can be rewritten, after some integrations by parts,  
as  a quadratic DHOST action of the form  
\bea\label{generalDHOST}
 S[ \tilde g_{\mu\nu},\phi] = \int d^4x \, \sqrt{- \tilde g} \left[ \tilde  f_0(\phi, \tilde  X)+ \tilde  f_1(\phi, \tilde  X) {\stackrel{\thicksim}{\Box}} \phi + \tilde  f_2(\phi,\tilde  X) \tilde  R + \sum_{A=1}^5 \tilde  a_A (\phi, \tilde  X) \tilde  L_A^{(2)} \right]\!\!,
\eea
with 
\be
\label{paramDHOST}
\begin{split}
\tilde  f_0  \; &= \; 3  \tilde  X^2 \, f_{2,\phi\phi} (\phi) \, , \quad 
\tilde  f_1  \; = \; 3 \tilde  X \, f_{2,\phi} (\phi) \, , \quad \tilde  f_2  =- \tilde  X{f}_2 (\phi) \,, \quad
\tilde  a_1  = a_1 (\phi)\, , \quad \tilde  a_2   = a_2  (\phi) 
\, , \\
\tilde  a_3  & = \frac{2}{\tilde  X}  \big[  a_1(\phi) + 2 a_2 (\phi) \big]\, , \quad
 \tilde  a_4   =  -\frac{2}{\tilde  X} \big[ 3 f_2(\phi) + a_1 (\phi)\big]  \, , \quad
\tilde a_5   = \frac{2 }{\tilde  X^2}  \big[ a_1 (\phi)+2 a_2 (\phi) \big]\, .
\end{split}
\ee

If $f_2=0$, one can easily check that the above theories belong to the subclass IIIa (or M-I), as defined in \cite{Achour:2016rkg} (or in \cite{Crisostomi:2016czh}). If $f_2\neq 0$ but $a_1+a_2=0$, then one gets theories belonging to the subclass Ia (or N-I).  In particular, this is  the case of the original mimetic theory. Finally, in the generic case where $f_2\neq 0$ and $a_1+a_2\neq 0$,  these theories are in the class II.
 More precisely, they belong to subclass IIa (or N-III) if $f_2 \neq a_1$ and to subclass IIb (or N-IV)  if $f_2 = a_1$.

The above calculation can be generalized to the case of a non-invertible disformal transformation, given by eq.~\eqref{disformal} with 
eq.~\eqref{non-invertibility}, 
using the results of \cite{Achour:2016rkg}.
Starting from the  action \eqref{quadmim}, 
one obtains an action of the form   \eqref{generalDHOST}, where now $\tilde{f}_2 $ is an arbitrary function and
\be
\label{mimetic_DHOST}
\begin{split}
\tilde{a}_1  &=  \tilde{c}_1(\phi)\left( \frac{\tilde{f}_2(\phi,\tilde X)}{\tilde X}\right)^3 + \frac{\tilde{f}_2(\phi,\tilde X)}{\tilde X} \, ,\qquad
\tilde{a}_2   =  \tilde{c}_2(\phi)\left( \frac{\tilde{f}_2(\phi,\tilde X)}{\tilde X}\right)^3 - \frac{\tilde{f}_2(\phi,\tilde X)}{\tilde X} \, ,
\\
\tilde a_3 & =  \frac{4 \tilde c_2 \tilde  f_2 ^2 \left(3 \tilde  X \tilde  f_{2,X}-2 \tilde  f_2 \right)-2 \tilde  c_1 \tilde f_2^2 \left( \tilde  f_2 -2 \tilde  X \tilde  f_{2,X}\right)+2 \tilde  X^2 \left(\tilde  f_2-2 \tilde X \tilde  f_{2,X}\right)}{\tilde X^4} \;, \\
\tilde a_4 &= -\frac{2 \tilde c_1 \tilde f_2^3}{\tilde X^4}-\frac{4 \tilde f_{2,\tilde X}}{\tilde X}+\frac{8 \tilde f_{2, \tilde X}^2}{\tilde f_2} \;,\\
\tilde a_5 & =  \frac{2 \tilde c_1 \tilde f_2^2 \left(3 \tilde f_2^2+6 \tilde X^2 \tilde f_{2,X}^2-8 \tilde f_2 \tilde X \tilde f_{2,X}\right)+4 \tilde c_2 \tilde f_2^2 \left(2 \tilde f_2-3 \tilde X \tilde f_{2,X}\right)^2- 2 \tilde X^2 \left(\tilde f_2-2 \tilde X \tilde f_{2,X}\right)^2}{\tilde f_2 \tilde X^5} \;,     
\end{split}
\ee
with $\tilde{c}_1(\phi)$ and $\tilde{c}_2(\phi)$  two independent functions of $\phi$ only.
For mimetic theories in the subclass Ia, one must impose the additional condition that $\tilde c_2=-\tilde c_1$. 
One can check that the above  functions satisfy the degeneracy conditions given in  \cite{Langlois:2015cwa}. 
The remaining functions, $\tilde{f}_0$ and $\tilde{f}_1$, have rather complicated expressions in terms of $\tilde{f}_2$ and  we do not  give them explicitly here. 

We can also extend our discussion to  cubic (or even higher) 
theories adding to \eqref{quadmim} all the  terms that are cubic 
in second derivatives, i.e.,
\bea
b_1(\phi)\chi_1^3 + b_2(\phi)\chi_1 \chi_2 + b_3(\phi)\chi_3 \, .
\eea
These correspond to the first three elementary cubic Lagrangians $L_1^{(3)}$, $L_2^{(3)}$ and $L_3^{(3)}$ introduced
in \cite{BenAchour:2016fzp}.  Performing the  non-invertible conformal transformation \eqref{conf_trans} leads  
to a minimal cubic DHOST for the metric $\tilde{g}_{\mu\nu}$, whose Lagrangian is
\bea
\sum_{A=1}^{10} \tilde{b}_A(\phi,\tilde{X}) \tilde{L}_A^{(3)} \;,
\eea
with
\bea
&&\tilde{b}_1 \, = \, -\frac{b_1}{\tilde X^3} \, , \quad
\tilde{b}_2 \, = \, -\frac{b_2}{\tilde X^3} \, , \quad
\tilde{b}_3 \, = \, -\frac{b_3}{\tilde X^3} \, , \quad
\tilde{b}_4 \, = \, -\frac{6b_1+2 b_2}{\tilde X^4} \, , \nonumber \\
&&\tilde{b}_5 \, = \, \frac{2b_2}{\tilde X^4} \, , \quad
\tilde{b}_6 \, = \, \frac{2b_2-3b_3}{\tilde X^4} \, , \quad
\tilde{b}_7 \, = \, \frac{3b_3}{\tilde X^4} \, , \quad
\tilde{b}_8 \, = \, \frac{3b_3 + 4 b_2}{\tilde X^5} \, , \nonumber \\
&&
\tilde{b}_9 \, = \, -3\frac{4b_1+2b_2+b_3}{\tilde X^5} \, , \quad
\tilde{b}_{10} \, = \, -2\frac{4b_1 + 2b_2+b_3}{\tilde X^6} \, .
\eea
According to the classification of cubic DHOST theories in \cite{BenAchour:2016fzp},  
these theories  either belong to  class ${}^3$M-I, if $9b_1+2b_2 \neq 0$, or to class ${}^3$M-III, if $9b_1+2b_2 = 0$, and are compatible with the quadratic theories (\ref{paramDHOST}).

\subsection{Lagrange multiplier formulation}
The mimetic action \eqref{mimetic} can also be reformulated 
as an action for the metric ${g}_{\mu\nu}$ instead of $\gt_{\mu\nu}$  as follows \cite{Barvinsky:2013mea}
\bea\label{lagrangeform}
{S'}[{g}_{\mu\nu},\phi]=\int d^4 x \, \sqrt{-g} \left[ {\cal L}( \phi,\phi_\mu, \phi_{\mu \nu}; {g}_{\mu\nu})
\, + \, \lambda({X} + 1) \right] \, ,
\eea
where  
$\lambda$ enforces the mimetic constraint ${X} =-1$, given in eq.~\eqref{Xuno}. If we express the action as
in \eqref{MimeticDHOST}, this action takes the form
\bea\label{lagform3}
{S'}[{g}_{\mu\nu},\phi]=\int d^4 x \, \sqrt{-g} \left[ {f}_2(\phi) {R}  \, + \, 
{\cal L}_\phi(\phi,\chi_1,\cdots,\chi_n) \, + \, \lambda({X} + 1) \right] \, .
\eea

Note that there is no ${X}$ dependence in this Lagrangian,  except in the term proportional to $\lambda$. 
If one assumes an explicit dependence of  ${f}_2$  and ${\cal L}_\phi$   on ${X}$
(as  in e.g. \cite{Arroja:2015wpa,Arroja:2015yvd}), as well as on $\vartheta_n = \phi^\mu\phi^\nu [{\phi}]_{\mu\nu}^n$, then the equations of motion  turn out  to be equivalent to those obtained from the   Lagrangian \eqref{lagform3} where ${X}$ is replaced by $- 1$ and all the $\vartheta_n$ by zero.

Let us illustrate this point  with a simple Lagrangian of the form
\bea
\label{example}
\int d^4 x \, \sqrt{-g} \left[ {f}_2(\phi,X) {R}  \, + \, 
{\cal L}_\phi(\phi, X,\vartheta_1) \, + \, \lambda({X} + 1) \right] \, ,
\eea
where we have introduced an explicit dependence on $X$ and $\vartheta_1=\phi^\mu\phi_{\mu\nu}\phi^\nu$. This action leads to the mimetic constraint $X=-1$ and to the 
metric equation of motion
\be
\begin{split}
\label{EQmimetic}
f_2 \, G_{\mu\nu}  = \ &  \left(\frac{1}{2} {\cal L}_\phi - \Box f_2 \right) g_{\mu\nu} + \nabla_\mu\nabla_\nu f_2  - {\cal L}_{\phi,\vartheta_1}
X_\mu \phi_\nu  \\
&  - \left[ \lambda + f_{2,X} R + {\cal L}_{\phi,X} + \nabla_\alpha (\phi^\alpha {\cal L}_{\phi,\vartheta_1}) \right] \phi_\mu \phi_\nu \, .
\end{split}
\ee
The equation for the scalar field $\phi$ can be obtained from the Bianchi identity. 
The trace of the above equation enables us  to express the Lagrange multiplier $\lambda$ in terms of the metric and the scalar field, namely
\bea
\lambda =  2 {\cal L}_\phi - 3 \Box f_2 + f_2 \, R -f_{2,X} R - {\cal L}_{\phi,X} - \nabla_\alpha (\phi^\alpha {\cal L}_{\phi,\vartheta_1}) \, ,
\eea 
where we have used the mimetic constraint and also $ X_\mu=0$.
Substituting this expression back into  (\ref{EQmimetic})  leads to the traceless 
metric 
equation
\bea
f_2 \, G_{\mu\nu} =  \left(\frac{1}{2} {\cal L}_\phi - \Box f_2 \right) g_{\mu\nu} + \nabla_\mu\nabla_\nu f_2 +
(2 {\cal L}_\phi - 3 \Box f_2 + f_2 \, R) \phi_\mu \phi_\nu\,.
\eea
 This is exactly the same equation as the one  obtained   from 
\eqref{example} with  $X=-1$ and $\vartheta_1=0$ directly in the Lagrangian. 
This conclusion remains valid when  ${\cal L}_\phi$ is an arbitrary function of the variables $\vartheta_n$.

\section{Effective approach to cosmological perturbations}
\label{Sec:EFT}

Since mimetic theories can be formulated in different ways, there are various but equivalent approaches to study cosmological  perturbations in mimetic gravity. For our purpose, it will be convenient to use the  unifying formulation  given by the effective approach developed in \cite{Gubitosi:2012hu,Gleyzes:2013ooa,Gleyzes:2014rba} and extended to higher-order scalar-tensor theories in 
\cite{Langlois:2017mxy}.

Let us start by briefly reviewing this  approach for  general DHOST theories \cite{Langlois:2017mxy}.
We will use the ADM parametrization for the metric:
\bea
ds^2 \; = \; -N^2 dt^2 + h_{ij} (dx^i + N^i dt) (dx^j+N^j dt) \, ,
\eea
where $N$ is the lapse function, $N^i$ the shift vector,  $h_{ij}$ the three dimensional metric and the   components of the extrinsic curvature tensor $K_{ij}$ are given by
\bea
K_{ij} \; = \; \frac{1}{2N} (\dot{h}_{ij} - D_i N_j - D_j N_i) \, , 
\eea
where $D_i$ denotes the covariant derivative compatible with $h_{ij}$.

We work in the so-called unitary gauge  where the
 constant time hypersurfaces coincide with the uniform scalar field hypersurfaces. 
 Assuming that the evolution of $\phi$ is monotonic, without loss of generality, we choose $\phi(t)= t$ on the background solution. 
In this gauge, we can  expand any action in terms of the metric and matter fluctuations up to quadratic order. Adopting the notation of \cite{Langlois:2017mxy} for the effective description of  DHOST theories, the gravitational part of the action 
expanded around a flat FLRW metric $ds^2 = - dt^2 + a^2(t)d \vec x^2$ can  always be written in the form 
\be
\begin{split}
\label{SBAction0}
& S_{\rm ETofDE} = \int d^3x \,  dt \,   \sqrt{h}  \frac{  M^2}2\bigg\{ \delta  K_{ij }\delta   K^{ij}- \left(1+\frac23 \aL\right)\delta   K^2  +(1+ \alphaT)  {}^{(3)}\!\R \\
&  +   H^2 \alphaK \delta   N^2+4   H  \alphaB \delta   K \delta   N+ ({1+ \alphaH})   {}^{(3)}\!\R  \delta   N   +  4  \bun  \delta   K  {\delta \dot  {  N} }   +  \bdeux  {\delta \dot {  N}^2} +  \frac{ \btrois}{  a^2}(\partial_i \delta   N )^2   
\bigg\} \; ,
\end{split}
\ee
where $\delta N$ and  $\delta  K_{ij}$ are respectively the perturbations of the lapse and of the extrinsic curvature, $\delta  K$ is the trace of $\delta  K_{ij}$ and $ {}^{(3)}\!R$ is the 3-dimensional Ricci scalar. 

In \cite{Langlois:2017mxy} it was shown that  the effective parameters introduced above satisfy either of the degeneracy conditions
\begin{align}
\label{Ia}
&\CI:\qquad \aL =0\,, \qquad \bdeux=-6\bun^2 \;,  \qquad  \btrois=-2\bun\left[2(1+\alphaH)+\bun (1+\alphaT) \right] \;, \\
\label{IIa}
&\CII:\qquad \bun=- (1+\aL)\frac{1+\alphaH}{1+\alphaT}\,, \qquad \bdeux=-6(1+\aL) \frac{(1+\alphaH)^2}{(1+\alphaT)^2} \;, \qquad \beta_3= 2 \frac{(1+\alphaH)^2}{1+\alphaT} \;,
\end{align}
depending on the class of the DHOST theory considered (see Table $1$ in App. B1 of \cite{Langlois:2017mxy}). 
As a consequence, this restricts by three  the number of  independent parameters. 

In the following, we use this approach to study  cosmological perturbations of mimetic theories, first  in Sec.~\ref{DHOSTform}, using the DHOST formulation  of mimetic actions and then in the
Lagrange multiplier formulation \eqref{lagform3} in Sec.~\ref{Lagrangemult}. The two formulations are equivalent and lead to the same results, as expected. 

For the analysis of linear perturbations,
we will also include matter in the action, described in terms of a  scalar field $\psi$ with a Lagrangian depending on its first derivatives \cite{ArmendarizPicon:2000dh},
\be
\label{matterLag}
S_{\rm m} =\int d^4 x \sqrt{- g}  \, P(Y) = \int d^4 x \sqrt{- \tilde g}\,  \tilde  X^2 P( \tilde  Y)  \;,
\ee
where
\be
Y\equiv g^{\mu \nu} \partial_\mu \psi \partial_\nu \psi  = 
 - \frac{1}{\tilde  X} \tilde  g^{\mu \nu} \partial_\mu \psi \partial_\nu \psi  \equiv \tilde  Y \;.
\ee
We have written the matter action explicitly  in terms of the physical metric $g_{\mu\nu}$ and in terms of the auxiliary metric $\tilde g_{\mu\nu}$, since we will need  both expressions to analyse linear perturbations in the DHOST formulation and in the Lagrange multiplier formulation.

\section{Perturbations in the DHOST formulation}
\label{DHOSTform}

In this section, we study  linear perturbations of mimetic theories in their DHOST formulation. In the first subsection we will characterize the effective formulation of all mimetic DHOST and thus we will consider the general non-invertible transformation \eqref{disformal}. 
It is sufficient to use the simplest non-invertible transformation \eqref{conf_trans} to study the linear perturbations 
of all mimetic theories, which we do in the subsequent subsection.

\subsection{Effective description}
\label{DHOSTUG}

In the unitary gauge, using \eqref{Xh}, one obtains
\bea
\frac{\dot \phi^2(t)}{N^2} \; = \; - X \; = \; - \frac{1}{h(t)} \, , 
\eea
which implies that $N$ has no spatial dependence. Moreover,  
one can always redefine the time and the scalar field such that $\phi = t$ and $N=1$ without loss of generality, see also App.~\ref{fieldredef}.
Furthermore, according to the definition of $\chi_n$ given in \eqref{chidef},   we find, in unitary gauge,
\bea\label{chin}
\chi_n \; = \; (-1)^n \text{tr}(K^n) \, \equiv \, (-1)^n K_{i_1}^{i_2} K_{i_2}^{i_3} \cdots K^{i_n}_{i_{n-1}}K_{i_n}^{i_1} \, ,
\eea
so that the action 
\eqref{MimeticDHOST} 
reduces to 
\bea\label{unitarygaugeaction}
S
=  \int d^4 x \, \sqrt{h} &&\!\!\! \!\!\! \!\!\! \Big[ f_2 {}^{(3)}\!R - 2 f_{2,\phi} K + f_2 (K_{ij} K^{ij} - K^2) 
 \nonumber \\
&& \!\!\! \! + \, {\cal L}_\phi\left(t,-K;\cdots,(-1)^p \text{tr}(K^p) \right)\Big] \, ,
\eea
where we have used the Gauss-Codazzi relation\footnote{The Gauss-Codazzi relation is 
\be
R = {}^{(3)}\! R + K_{\mu \nu} K^{\mu \nu} - K^2 + 2 \nabla_\nu (n^\nu K - n^\mu \nabla_\mu n^\nu) \;.
\ee} and integrated by parts its last term.

We need to expand this action up to quadratic order in the perturbations. 
The expansion of ${\cal L}_\phi$  is given by
\bea
\label{foursix}
{\cal L}_\phi  =  \bar {\cal L}_\phi (t) + c_0(t)  \delta K + c_1 (t) \delta K_i^j \delta K_j^i +
c_2 (t) (\delta K)^2 + {\cal O}(\delta K_{ij}^3) \, ,
\eea
where $ \bar {\cal L}_\phi (t)$ denotes  ${\cal L}_\phi$ evaluated on the background. 
In the second term, $c_0$ is defined by
\be
c_0(t) \equiv \sum_{n=1} (-1)^n n H^{n-1} {\cal L}_{\phi,n}  \;, 
\ee
where ${\cal L}_{\phi,n}$ stands for the derivative of ${\cal L}_\phi$ with respect to $\chi_n$ evaluated on the background.
The coefficients $c_1$ and $c_2$ in the third and fourth terms of \eqref{foursix} are given by  combinations of the first and second derivatives of ${\cal L}_\phi$ with
respect to $\chi_n$, respectively
\begin{align}
c_1 (t) & \equiv  \frac{1}{2} \sum_{n=1} (-1)^n n(n-1) H^{n-2} {\cal L}_{\phi,n} \, , \label{a1}\\
c_2 (t) & \equiv  \frac{1}{2} \sum_{(n,m)} nm H^{n+m-2} {\cal L}_{\phi,nm} \, , \label{a2}
\end{align}
where the right-hand sides are evaluated on the background.
Notice that in the simple case \eqref{quadmim}, these expressions reduce to $c_1=a_1$ and
$c_2=a_2$.

Using these results and integrating by parts the term proportional to $c_0(t)$, we can rewrite the quadratic action  in terms of the effective parameters introduced in the previous section as  
\be
\label{UGaction}
  S_{\rm ETofDE} = \int d^3x \,  dt \,  \sqrt{  h}  \frac{ M^2}2\bigg\{ \delta  K_{ij }\delta   K^{ij}- \left(1+\frac23 \aL\right)\delta  K^2  +(1+ \alphaT)  {}^{(3)}\! R   
\bigg\} \; ,
\ee
where 
\be
\label{MLT}
\frac{M^2}{2} = f_2 + c_1  \;, \qquad \aL =   - \frac{3}{2} \frac{c_1 +c_2}{ f_2 + c_1} \;, \qquad \alphaT =  - \frac{c_1}{f_2 + c_1} \;.
\ee
The parameters that do not appear 
in this action
(because $\delta N=0$), i.e. $ \alpha_{\rm K}$, $ \alpha_{\rm B}$, $ \alpha_{\rm H}$, $ \bun$, $  \bdeux$ and $ \btrois$, simply remain  undetermined.

Now, we can express this effective action in terms of the 
new metric $\tilde g_{\mu\nu}$, introduced in  \eqref{disformal}. The transformations of the parameters under general  disformal transformations have been given in \cite{Langlois:2017mxy} (see also \cite{Gleyzes:2015pma,DAmico:2016ntq} for earlier work) and we report them for convenience in App.~\ref{App:alphatrans}.
As shown in \cite{Langlois:2017mxy}, the structure of the action \eqref{SBAction0} is preserved under this  
transformation  of the metric. 
Therefore, the action takes the form
\be
\begin{split}
\label{SBActiontilde}
&\tilde  S_{\rm ETofDE} = \int d^3x \,  d \tilde t \,  \sqrt{ \tilde h}  \frac{\tilde M^2}2\bigg\{ \delta \tilde K_{ij }\delta  \tilde K^{ij}- \left(1+\frac23 \tilde\alpha_{\rm L} \right)\delta \tilde K^2  +(1+\tilde \alpha_{\rm T})  {}^{(3)}\!\tilde R  \\
&  + \tilde H^2\tilde\alpha_{\rm K} \delta \tilde N^2+4 \tilde H \tilde\alpha_{\rm B} \delta \tilde K \delta \tilde N+ ({1+\tilde\alpha_{\rm H}}) {}^{(3)}\! \tilde \R  \delta \tilde N   +  4 \tilde\bun  \delta \tilde K  {\delta \dot  {\tilde N} }   + \tilde\bdeux  {\delta \dot {\tilde N}^2} +  \frac{\tilde\btrois}{\tilde a^2}(\partial_i \delta \tilde N )^2   
\bigg\} \; ,
\end{split}
\ee
where the time-dependent parameters are related to $M^2$, $\aL$, $\alphaT$, via the non-invertible disformal transformation \eqref{disformal}.

\begin{figure}
\begin{center}
\includegraphics[width=10cm]{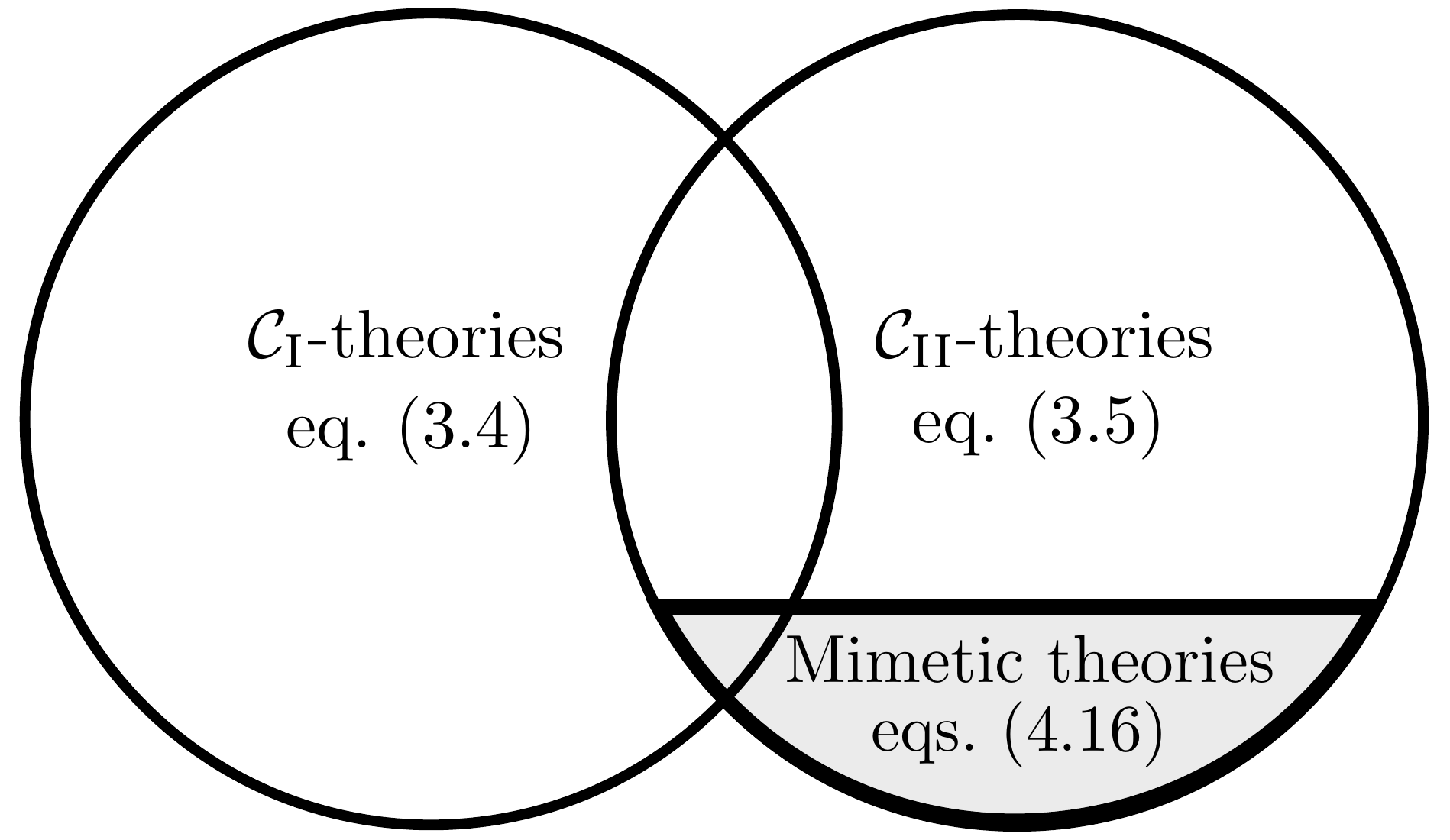}   \\
\end{center}
\caption{Theories of 
category 
$\CI$ and $\CII$ are respectively characterized by the conditions \eqref{Ia} and  \eqref{IIa} 
(with tildes on all coefficients).
 In this figure, they are respectively represented by the left- and right-hand side disc.  Note that some theories can be both $\CI$ and $\CII$, as shown by the overlapping region between the two discs. Mimetic theories are a subset of theories of 
  category $\CII$, verifying the conditions on $\tilde \alpha_{\rm K}$ and $\tilde \alpha_{\rm B}$ given by eq.~\eqref{alphaKB}. They are represented by the gray region. 
Each category, $\CI$, $\CII$ and mimetic, is preserved under invertible disformal transformations.
} \label{Plot}
\end{figure}

In particular, 
we define
\be
\label{defalphas2sec4}
 \tilde \alpha_{\rm Y} \equiv - \frac{\tilde X}{\tilde  A} \frac{\partial \tilde A}{\partial \tilde X} \, , \qquad  \tilde \alpha_{\rm D} \equiv - \frac{\tilde B}{\tilde  B+\tilde  A/\tilde  X} \; , \qquad \tilde \alpha_{\rm X} \equiv- \frac{\tilde X^2}{\tilde A}\frac{\partial \tilde B}{\partial \tilde X} \;,
\ee
with the property that 
\be
1+ \tilde \alpha_{\rm X} + \tilde \alpha_{\rm Y} =0 \;,
\ee
which follows from the non-invertibility, see eq.~\eqref{non-invertibility}.
One finds (see App.~\ref{App:alphatrans} for details)
\be 
\tilde {M}^2 ={ M^2 }{ \tilde A \sqrt{1+\tilde \alpha_{\rm D}} }\; , \qquad  \tilde{\alpha}_{\rm L} = \aL  \; , \qquad 
\tilde \alpha_{\rm T} = -1 + \frac{1 + {\alpha}_{\rm T}}{1 +  \tilde{\alpha}_{\rm D} } \; , 
\ee
while the other parameters can be expressed as 
\be 
\label{alphatilde2sec4}
\begin{split}
 \tilde{\alpha}_{\rm H}  &= -1 +  \tilde{\alpha}_{\rm Y} (1+\tilde{\alpha}_{\rm T}) \;,\\
 \tilde \bun &= - \tilde{\alpha}_{\rm Y}(1+\tilde \alpha_{\rm L}) \; ,\\
\tilde \bdeux & = - 6 \tilde \alpha_{\rm Y} ^2 (1+\tilde \alpha_{\rm L}) \; , \\
 \tilde \btrois & = 2  \tilde \alpha_{\rm Y}^2 (1+\tilde \alpha_{\rm T})   \; , 
\end{split}
\ee
and
\be
\begin{split}
\label{alphaKB} 
  \tilde{\alpha}_{\rm B}  & =  -(1+\tilde{\alpha}_{\rm L}) \left(1+\frac{\dot {\tilde{\alpha}}_{\rm Y}}{\tilde H} \right)     \;, \\
 { \tilde{\alpha}_{\rm K}} &=  6 \tilde{\alpha}_{\rm B}   -   \frac{6    }{ \tilde{M}^2 \tilde{a}^3 \tilde{H}^2} \frac{d}{d \tilde t} \Big( \tilde{M}^2 \tilde{a}^3 \tilde  H    \tilde{\alpha}_{\rm Y} \tilde{\alpha}_{\rm B}   \Big) \;.
\end{split}
\ee
Note that $\tilde{\alpha}_{\rm Y}$ can be expressed in terms of $\tilde{\alpha}_{\rm T}$ and $ \tilde{\alpha}_{\rm H}$ using the first equation in \eqref{alphatilde2sec4} and then substituted in the other expressions. As a consequence, from the point of view of the tilde frame, we find that $\tilde M^2$, $\tilde{\alpha}_{\rm L}$, 
$\tilde \alpha_{\rm T}$ and $\tilde \alpha_{\rm H}$ are free functions whereas the functions $\tilde\bun$, $\tilde\bdeux$ and $\tilde\btrois$ are determined, in terms of $\tilde{\alpha}_{\rm L}$, $\tilde{\alpha}_{\rm T}$ and $ \tilde{\alpha}_{\rm H}$, by conditions that coincide with the degeneracy conditions $\CII$, see eq.~\eqref{IIa}. Moreover,
$\tilde \alpha_{\rm K} $ and $\tilde \alpha_{\rm B}$  are fixed by the relations 
\eqref{alphaKB}.
Therefore, mimetic theories are 
particular  DHOST theories 
that
satisfy 
$\CII$  and eq.~\eqref{alphaKB}. For $\tilde \alpha_{\rm L}=0$ they satisfy both $\CI$, see eq.~\eqref{Ia}, and $\CII$.
 One can check that these conditions are preserved by invertible disformal transformations. 
See Fig.~\ref{Plot} for a summary on the classification of theories based on linear perturbations.

To illustrate these results, let us consider the simple case \eqref{conf_trans} where $\tilde{A}=-\tilde{X}$ and $\tilde{B}=0$. In this case, $\tilde \alpha_{\rm D} =0$ and $\tilde \alpha_{\rm Y} =-1$,
and we find that the parameters  \eqref{MLT} are transformed according to
\be
\tilde M^2 =   - 2 \tilde X (f_2 + c_1)  \;, \qquad \tilde \alpha_{\rm L} = - \frac{3}{2} \frac{c_1 +c_2}{ f_2 + c_1} \;, \qquad \tilde \alpha_{\rm T} =   - \frac{c_1}{f_2 + c_1} \;.
\ee
The remaining parameters can be written in terms of the three parameters above by using eq.~\eqref{alphatilde2sec4} with $\tilde \alpha_{\rm Y} =-1$.
Hence, in this special case $\tilde \alpha_{\rm H}$ is no 
longer
independent. 

\subsection{Linear perturbations}  
\label{Stability_DHOST}

Let us study the linear  
perturbations in mimetic theories in the presence of matter minimally coupled to $g_{\mu \nu}$ (see eq.~\eqref{matterLag}). To do so, it is sufficient to take a particular DHOST representative by choosing the special case of $\tilde A = - \tilde X$ and $\tilde B=0$.
Therefore, we consider eq.~\eqref{SBActiontilde} with the relations \eqref{alphatilde2sec4}
and  $\tilde \alpha_{\rm Y} =-1$.

Including matter and specializing the action \eqref{SBActiontilde} to scalar perturbations, defined by 
\begin{align}
\tilde h_{ij}  = \tilde a^2 (t)  e^{2 \tilde \zeta  } \delta_{ij} \;, \qquad \tilde N^i = \delta^{ij} \partial_j \tilde \chi  \;, \qquad
 \psi  = \psi_0(t) + \delta \psi   \;,
\end{align}
we obtain the total quadratic action\footnote{Note that, due to the definition of $X$, $P'$ has sign opposite  to $P$, so that the matter action has the correct sign in the action \eqref{SBActionfil}.}
\be
\begin{split}
\label{SBActionfil}
 S^\quadac = &\int d^3x \,  dt \, \tilde  a^3  \bigg\{ \frac{\tilde M^2}2 \bigg[ - 6 (1+ \tilde \alpha_{\rm L}) \dot \zeta^2 + \frac{2}{\tilde a^2} (1+\tilde \alpha_{\rm T}) (\partial_i \zeta)^2 + \big[4 (1+\tilde \alpha_{\rm L}) \dot \zeta 
\\
& -2 P' \dot \psi_0 \delta \psi \big] \partial^2 \tilde \chi 
  - \frac23 \tilde \alpha_{\rm L} (\partial^2 \tilde \chi )^2  \bigg] - \frac{P'}{c_{\rm m}^2} \left[ \delta \dot \psi^2 - \frac{c_{\rm m}^2}{\tilde a^2} (\partial_i \delta \psi)^2 \right] - 6  P' \dot \psi_0\,  \delta \dot\psi \, \zeta \bigg\}
 \; ,
\end{split}
\ee
where we have introduced  the  sound speed of matter fluctuations \cite{Garriga:1999vw}
\be
c_{\rm m}^2 \equiv \frac{P'}{P' - 2 \dot \psi_0^2 P''} \;,
\ee
and defined 
\be
\label{zetadef}
\zeta \equiv \tilde \zeta - \delta \tilde N \;.
\ee
Moreover,  note that $\delta \tilde N $ only appears  in the combination  \eqref{zetadef}. This is due
to the conformal invariance of the theory. Indeed, 
up to linear order,  the  metric can be written as 
\be
ds^2 
\simeq
 e^{2 \delta \tilde N} \Big[ - dt^2 + 2 N_i dx^i dt + \tilde a^2 e^{2 ( \tilde \zeta - \delta \tilde N) } d \vec x^2 \Big] \;,
\ee
which shows that, for a conformally invariant theory, only the combination given by eq.~\eqref{zetadef} matters.

Following the analysis of \cite{Langlois:2017mxy}, we distinguish two cases. For $\tilde \alpha_{\rm L} \neq 0 $, variation  with respect to $\tilde \chi$ yields
\be
\partial^2 \tilde \chi = \frac{3}{\tilde \alpha_{\rm L}} \left[ (1+\tilde \alpha_{\rm L})\dot \zeta -\frac{P'}2 \dot \psi_0 \delta \psi \right]\;,
\ee 
which can be plugged back into the action to give
\be
\begin{split}
\label{SBActionfil2}
 S^\quadac = \int d^3x \,  dt \, \tilde  a^3  \bigg\{& A \bigg[     \dot \zeta^2 - c_s^2 \frac{(\partial_i \zeta)^2}{\tilde a^2}   - P' \dot \psi_0 \delta \psi \dot \zeta   \bigg]
- 6 P' \dot \psi_0 \delta \dot \psi \, \zeta \\
&   - \frac{P'}{c_{\rm m}^2} \left[ \delta \dot \psi^2 - \frac{c_{\rm m}^2}{\tilde a^2} (\partial_i \delta \psi)^2 \right]  + \frac34 \frac{\tilde M^2 P'{}^2 \dot \psi_0^2 }{\tilde \alpha_{\rm L}}  \delta \psi^2  \bigg\}
 \; ,
\end{split}
\ee
with
\be
A \equiv \frac{3 \tilde M^2}{\tilde \alpha_{\rm L}} (1+\tilde \alpha_{\rm L}) \;, \qquad c_s^2 \equiv  -\frac{\tilde \alpha_{\rm L}( 1+\tilde \alpha_{\rm T})}{3 (1+\tilde \alpha_{\rm L})} \;.
\ee

At this stage, we remind the reader that the normalization of the quadratic action for tensor modes is fixed by $\tilde M^2/8$ and that the speed of propagation of tensors is given by 
$c_T^2 = 1+\tilde\alpha_{\rm T}$
 in this formulation \cite{Gleyzes:2013ooa,Gleyzes:2014rba}.
Since  $(1+ \tilde \alpha_{\rm L})/{\tilde \alpha_{\rm L}}>0$ to avoid that $\zeta$ propagates a ghost,  the propagation speed squared of scalar fluctuations  has a sign opposite to that of tensor fluctuations. This implies an instability either in the scalar or in the tensor sector. 
In fact, this is true
not only for mimetic gravity but  for any DHOST theory 
in the category ${\cal C}_{\rm II}$ defined in  \cite{Langlois:2017mxy}.

For $\tilde \alpha_{\rm L} = 0$, the analysis is different. The variation of the action \eqref{SBActionfil} with respect to $\tilde \chi$ implies a relation between $\dot \zeta$ and $\delta \psi$,
i.e.,
\be
\delta \psi  = \frac{2  \dot \zeta}{P' \dot \psi_0} \;.
\ee
This can be used to replace $\delta \psi$ in the action \eqref{SBActionfil2} above, to yield an action for $\zeta$ only. 
This action contains a $\ddot \zeta^2$ term, which leads to a fourth-order equation of motion. This is consistent with the results of \cite{Takahashi:2017pje,Ganz:2018mqi}, where one can find discussions on their  physical interpretation. 
Notice that, together with the degeneracy condition $\CII$, the mimetic condition eq.~\eqref{alphaKB} is crucial for this result. 

Let us finish with a short discussion on the case $\tilde \alpha_{\rm L}=0$ without matter. In that case, one must go back to the action \eqref{SBActionfil}, which reduces to
\bea\label{Squadlam}
S^{\rm quad} = \int d^4x \, \tilde  a^3 \frac{\tilde M^2}{2} \left\{ -6 \dot\zeta^2 + \frac{2(1+\tilde \alpha_{\rm T})}{a^2} (\partial_i \zeta)^2 + 4\dot{\zeta}\Delta\tilde  \chi \right\} \, .
\eea
From this action, $\zeta$ and $\tilde \chi$ satisfy the equations
\bea
\dot{\zeta}\; = \; 0 \, , \qquad
\frac{d}{dt} \left( \tilde a^3 \tilde M^2 \tilde \chi \right) + \tilde a \tilde M^2 (1+\tilde \alpha_T) \zeta \; = \; 0 \, ,
\eea
which shows that $\zeta$ is not a dynamical variable. Replacing the second equation into the first one we find a second-order equation for $\tilde \chi$, 
\bea\label{psieq}
\frac{d}{dt} \left[ \frac{1}{ \tilde a  \tilde M^2(1+ \tilde \alpha_{\rm T})} \frac{d}{dt} \left(  \tilde a^3 \tilde M^2 \tilde \chi \right)\right] \; = \; 0 \, ,
\eea
which involves no gradient. Thus,  in the absence of matter $\tilde \chi$ behaves as a scalar with a vanishing speed of sound, as it was found in \cite{Mukohyama:2009mz,Lim:2010yk,Chamseddine:2013kea} and more generally in \cite{Arroja:2015yvd}.

\section{Perturbations in the Lagrange multiplier formulation}
\label{Lagrangemult}
We  now discuss  
linear perturbations in the Lagrange multiplier  formulation, with the action
\bea\label{quadraticaction}
{S}[{g}_{\mu\nu},\phi]=\int d^4 x \, \sqrt{{g}} \left[ {f}_2(\phi) {R}  \, + \, {\cal L}_\phi(\phi,\chi_1,\cdots,\chi_n)  + \lambda(X+1)\right] \,.
\eea
In unitary gauge, the equation for $\lambda$ is no longer an equation of motion but reduces to a constraint that fixes the lapse to $N=1$. Hence, from the beginning, we can set $N=1$ in the action.
The rest of the analysis is identical to what  is done at the beginning of Sec.~\eqref{DHOSTUG} and 
it is  straightforward to get eq.~\eqref{UGaction}.

Let us now compute the quadratic action for linear perturbations, starting directly from eq.~\eqref{UGaction}.
We can concentrate on the scalar perturbations and, analogously to what was
done in the previous section, we introduce the variables $\zeta$ and  $\chi$, but this time without a tilde,
\be
h_{ij}=a^2 (t) \, e^{2\zeta} \delta_{ij} \;, \qquad N^i=\delta^{ij} \, \partial_j \chi  \, .
\ee
Substituting into the quadratic action \eqref{UGaction}, we obtain an action for the perturbations $\zeta$ and $\chi$, 
\be
\begin{split}
\label{SBActionfil3}
 S^\quadac = \int d^3x \,  dt \,   a^3  \bigg\{& \frac{  M^2}2 \bigg[ - 6 (1+   \alpha_{\rm L}) \dot \zeta^2 + \frac{2}{  a^2} (1+  \alpha_{\rm T}) (\partial_i \zeta)^2 + \bigg[4 (1+  \alpha_{\rm L}) \dot \zeta -\frac{4}{M^2} P' \dot \psi_0 \delta \psi \bigg] \partial^2   \chi 
\\
&  - \frac23   \alpha_{\rm L} (\partial^2   \chi )^2  \bigg] - \frac{P'}{c_{\rm m}^2} \left[ \delta \dot \psi^2 - \frac{c_{\rm m}^2}{  a^2} (\partial_i \delta \psi)^2 \right] - 6  P' \dot \psi_0\,  \delta \dot\psi \, \zeta \bigg\}
 \; .
\end{split}
\ee
This is the same action as eq.~\eqref{SBActionfil} but the tildes are absent from all quantities, not only from $\zeta$. 
Indeed, this action can be  obtained from eq.~\eqref{SBActionfil} by considering the relation between the metric perturbations in the two frames. From eq.~\eqref{conf_trans} one gets 
\be
N^i = \tilde N^i \;, \qquad h_{ij} = \frac{1}{\tilde N^2} \tilde h_{ij} \;, 
\ee
which yields 
\be
a=\tilde a\;, \qquad \chi=\tilde \chi\;, \qquad  \zeta = \tilde \zeta - \delta \tilde N\;. 
\ee
Notice that this last relation is the same as eq.~\eqref{zetadef}. 
From this action, it is obvious that the analysis is completely analogous to the one made in Sec.~\ref{Stability_DHOST}. A similar analysis in the formulation with the Lagrange multiplier has been performed in \cite{Takahashi:2017pje}.

\section{Conclusion} 
\label{sec6}

We have studied mimetic theories in the framework of Degenerate Higher-Order Scalar-Tensor (DHOST) theories. Indeed, as explained in Sec.~\ref{sec2}, mimetic theories can be viewed as 
particular DHOST theories characterized by an extra symmetry. 
In general, this extra symmetry is an invariance under a combination of conformal and disformal transformations of the auxiliary metric used for the variation of the action. 
From the Hamiltonian point of view, this symmetry  gives  a  constraint in the theory 
(in addition to the usual Hamiltonian and momentum constraints associated with diffeomorphism invariance), which is first class,
in contrasts with non-mimetic DHOST theories  characterized by a pair of extra second-class constraints. 

We have found that, generically (when $f_2 \neq 0$ and $a_1+a_2\neq 0$), mimetic theories belong to the class II of DHOST theories. However, some mimetic theories exist in the subclass Ia of DHOST theories (when $a_1+a_2= 0$), the subclass that also contains  Horndeski theories. For mimetic DHOST  theories, the six coefficients of the quadratic terms (i.e. $f_2$ and the five $a_i$) are specified in terms of a single function of $X$, e.g. $f_2(X,\phi)$ and two arbitrary functions of $\phi$ only ($c_1(\phi)$ and $c_2(\phi)$, which reduce to only one in the subclass Ia), as given by the expressions (\ref{mimetic_DHOST}).

In the second part of this work,  
we have investigated the  
linear perturbations around a FLRW solution in these theories, using the effective theory of dark energy, reviewed in Sec.~\ref{Sec:EFT}. 
We have applied this approach to two different formulations of mimetic theories. The first formulation, called ``DHOST'' formulation, is worked out in Sec.~\ref{DHOSTform}. The quadratic action is obtained starting from a generic DHOST theory,  using a non-invertible conformal metric transformation, eq.~\eqref{conf_trans}. The effect of this transformation on the parameters of the 
quadratic
action was derived in \cite{Langlois:2017mxy} and is reviewed in App.~\ref{App:alphatrans}. Using these results, the final quadratic action is given in terms of four independent parameters (instead of 6 independent parameters in the non-mimetic case), with the others fixed by the non-invertible   transformation. Interestingly, the quadratic parameters for mimetic theories always satisfy the condition $\CII$, defined in (\ref{IIa}). The few mimetic theories that belong to the subclass Ia satisfy both the conditions $\CI$ and $\CII$. This is illustrated in Fig. \ref{Plot}.

We have studied the quadratic action of mimetic theories in the presence of  external matter, which for simplicity we have taken in the form of a scalar field with Lagrangian dependent on its first derivatives. 
For $\tilde \alpha_{\rm L} \neq 0$ we have found an instability either in the scalar or in the tensor sector. In the case $\tilde \alpha_{\rm L} = 0$, which contains the original mimetic theory \cite{Chamseddine:2013kea}, 
we have shown that the dynamics of scalar  perturbations can be expressed in terms of an action for $\zeta$ only, which is quadratic in $\ddot \zeta$.
The same conclusion can be reached in the Lagrange multiplier formulation, as shown  in Sec.~\ref{Lagrangemult} and in Ref.~\cite{Takahashi:2017pje}.

\subsection*{Acknowledgements}
We thank  Shinji Mukohyama and Kazufumi Takahashi  for  instructive  discussions. D.L.~and K.N.~acknowledge support from the
call DEFI InFIniTI 2016/2017 of the CNRS, France.  F.V.~acknowledges financial support from ``Programme National de Cosmologie and Galaxies'' (PNCG) of CNRS/INSU, France and  the French Agence Nationale de la Recherche under Grant ANR-12-BS05-0002.

\appendix
\section{Fields transformations: useful formulae}

\subsection{Scalar field redefinition}
\label{fieldredef}
\noindent
Let us show that we can fix $h(\phi)$ to the constant values $\pm 1$ simply by a field redefinition in \eqref{non-invertibility}. 
Indeed,  if we assume that there exist a new field $\hat{\phi}$ and a one-variable function $F$ such that $\phi=F(\hat{\phi})$, then 
the disformal transformation \eqref{disformal} becomes
\bea
{g}_{\mu\nu} \; = \; \hat{A}(\hat\phi,\hat{X}) \tilde g_{\mu\nu} \, + \, \hat{B}(\hat\phi,\hat X) \hat\phi_\mu \hat\phi_\nu \, ,
\eea
where
\bea
\hat{X}=\frac{\tilde X}{(F')^2}\, , \quad
\hat{A}(\hat\phi,\hat{X})=\tilde A[F(\hat{\phi}),(F')^2 \hat{X}] \, , \quad
\hat{B}(\hat\phi,\hat{X})=(F')^2 \, \tilde B[F(\hat{\phi}),(F')^2 \hat{X}]\, .
\eea
Hence, the non-invertibility condition of the disformal transformation \eqref{non-invertibility} becomes 
\bea
\frac{\hat{A}(\hat\phi,\hat{X})}{\hat X} + \hat{B}(\hat\phi,\hat{X}) \; = \; \hat{h}(\hat\phi) \quad
\text{with} \quad
\hat h({\hat \phi}) \equiv {(F')^2} \, {h[F(\hat{\phi})]} \, .
\eea
Now, if we assume that the sign of $h(\phi)$ is constant, we can always find a redefinition of the scalar field which allows
to fix the function $\hat{h}$ to $\pm$ 1. Indeed, this is the case if $F$ satisfies the differential equation
\bea
 \frac{dF}{dx} \; = \; \frac{1}{\sqrt{\vert h[F(x)] \vert}} \, ,
\eea 
which can be (at least formally) easily  integrated according to
\bea
F(x) \; = \; G^{-1}(x+c) \, , \qquad
G(y) \equiv \int dy \, {\sqrt{\vert h(y) \vert}} \, ,
\eea
where $c$ is a constant.

\subsection{Disformal transformations of the metric}
\label{disftran}
Let us show that one can fix the functions $\tilde A$ and $\tilde B$ to $\tilde A=\pm \tilde X$ and $\tilde B=0$ by a redefinition of the metric $\tilde g_{\mu\nu}$
using a disformal transformation. For that, we are looking for an invertible disformal transformation of the metric $\tilde g_{\mu\nu}$
\bea\label{dis2}
\tilde g_{\mu\nu} \; = \; C(\phi,\hat{X}) \hat{g}_{\mu\nu} + D(\phi,\hat{X}) \phi_\mu \phi_\nu 
\eea
such that
\bea
{g}_{\mu\nu} \; = \; \pm \hat{X} \, \hat{g}_{\mu\nu} \, .
\eea 
This condition is possible if the functions $C$ and $D$ satisfy the conditions
 \bea
 \tilde A(\phi,\tilde X) \, D(\phi,\hat{X}) + \tilde B(\phi,\tilde X) =  0 \, ,\qquad 
 \tilde A(\phi,\tilde X) \, C(\phi,\hat{X}) \; = \; \hat{X} \, ,
 \eea
 with
 \bea\label{XhatX}
 \tilde X = \frac{\hat{X}}{C(\phi,\hat{X})  + \hat{X}D(\phi,\hat{X}) } \, .
 \eea
These conditions can be satisfied (at least locally) when the disformal transformation \eqref{dis2} is invertible,
in which case the relation \eqref{XhatX} is well-defined.

\subsection{Disformal transformation in the effective description}
\label{App:alphatrans}
Let us consider the transformation
\be
\label{gCgtilde}
 g_{\mu \nu} = \tilde A( \phi,\tilde X) \tilde g_{\mu \nu} + \tilde B (\phi,\tilde X) \partial_\mu \phi \partial_\nu \phi \;.
\ee
Following \cite{Langlois:2017mxy}, we review here the effect of this  disformal transformation on the parameters $M^2$, $\alphaK$, $\alphaB$,
$\alphaT$, $\alphaH$, $\aL$, $\bun$, $\bdeux$ and $\btrois$. 
Let us  define
\be
\label{defalphas2}
 \tilde \alpha_{\rm Y} \equiv - \frac{\tilde X}{\tilde  A} \frac{\partial \tilde A}{\partial \tilde X} \, , \qquad  \tilde \alpha_{\rm D} \equiv - \frac{\tilde B}{\tilde  B+\tilde  A/\tilde  X} \; , \qquad \tilde \alpha_{\rm X} \equiv- \frac{\tilde X^2}{\tilde A}\frac{\partial \tilde B}{\partial \tilde X} \;.
\ee
The effective parameters in the quadratic action derived from ${S}$ and  those associated with $\tilde S$ are related via the transformations  given in eq.~(2.22) of Ref.~\cite{Langlois:2017mxy}. One finds
\be 
{M}^2 =\frac{\tilde M^2 }{ \tilde A \sqrt{1+\tilde \alpha_{\rm D}} }\; , \qquad \aL = \tilde{\alpha}_{\rm L} \; , \qquad 
1 + {\alpha}_{\rm T}=  (1+\tilde \alpha_{\rm D} ) (1 +  \tilde{\alpha}_{\rm T} )\; , 
\ee
and
\be 
\label{alphatilde}
\begin{split}
({1+\tilde{\alpha}_{\rm X}+\tilde{\alpha}_{\rm Y}} ) (1+ {\alpha}_{\rm H}) &=   1+\tilde{\alpha}_{\rm H}-\tilde{\alpha}_{\rm Y} (1+\tilde{\alpha}_{\rm T}) \;,\\
(1+ \tilde \alpha_{\rm D} ) ({1+\tilde{\alpha}_{\rm X}+\tilde{\alpha}_{\rm Y}} )  {\bun}&=  \tilde \bun+  \tilde{\alpha}_{\rm Y}(1+\tilde \alpha_{\rm L}) \; ,\\
(1+ \tilde \alpha_{\rm D} )^2 (1+\tilde{\alpha}_{\rm X}+\tilde{\alpha}_{\rm Y})^2 {\bdeux}&=  \tilde \bdeux-6\tilde{\alpha}_{\rm Y}(\tilde{\alpha}_{\rm Y} (1+\tilde \alpha_{\rm L} )+2\tilde \bun) \; , \\
(1+ \tilde \alpha_{\rm D} )(1+\tilde{\alpha}_{\rm X}+\tilde{\alpha}_{\rm Y})^2 {\btrois}&=   \tilde \btrois+2 \tilde{\alpha}_{\rm Y}^2(1+\tilde{\alpha}_{\rm T})-4\tilde{\alpha}_{\rm Y}(1+\tilde{\alpha}_{\rm H})  \; .
\end{split}
\ee
We are interested in a non-invertible transformation \eqref{gCgtilde}, where $1+ \tilde \alpha_{\rm X} + \tilde \alpha_{\rm Y} =0 $. Therefore, the left-hand side of eq.~\eqref{alphatilde}  vanishes so that one has
\be 
\label{alphatilde2}
\begin{split}
 1+\tilde{\alpha}_{\rm H}  &=  \tilde{\alpha}_{\rm Y} (1+\tilde{\alpha}_{\rm T}) \;,\\
 \tilde \bun &= - \tilde{\alpha}_{\rm Y}(1+\tilde \alpha_{\rm L}) \; ,\\
\tilde \bdeux & = - 6 \tilde \alpha_{\rm Y} ^2 (1+\tilde \alpha_{\rm L}) \; , \\
 \tilde \btrois & = 2  \tilde \alpha_{\rm Y}^2 (1+\tilde \alpha_{\rm T})   \; .
\end{split}
\ee
Note that the parameter $\tilde{\alpha}_{\rm Y} $ can be expressed in terms of $\tilde{\alpha}_{\rm H}$ and $\tilde{\alpha}_{\rm T}$ using the first of these equations and replaced in the other three equations, reproducing the degeneracy conditions $\CII$, eq.~\eqref{IIa}. Therefore, mimetic theories  satisfy the conditions $\CII$.

Using the above relations, from eq.~(C.14) of Ref.~\cite{Langlois:2017mxy} one also finds, for $1+ \tilde \alpha_{\rm X} + \tilde \alpha_{\rm Y} =0 $,
\be 
\label{alphatilde3}
\begin{split}
 \tilde{\alpha}_{\rm B}  & =  -(1+\tilde{\alpha}_{\rm L})  \left( 1 + \frac{\dot {\tilde \alpha}_{\rm Y}}{\tilde H} \right)   \;, \\
 { \tilde{\alpha}_{\rm K}} &= 6  \tilde{\alpha}_{\rm B}   -   \frac{6    }{ \tilde{M}^2 \tilde{a}^3 \tilde{H}^2} \frac{d}{d \tilde t} \Big( \tilde{M}^2 \tilde{a}^3 \tilde  H    \tilde{\alpha}_{\rm Y}  \tilde{\alpha}_{\rm B}  \Big) \;.
\end{split}
\ee
We can thus conclude that mimetic theories are 
particular DHOST theories which satisfy $\CII$ {\em and} the above relations. 
One can check that mimetic theories are closed under invertible conformal/disformal transformations.

\bibliographystyle{utphys}
\bibliography{EFTDE}

\end{document}